\documentclass[twocolumn,amsmath,amssymb,prb,superscriptaddress]{revtex4}

\pdfoutput=1

\usepackage[pdftex]{graphicx}
\usepackage[latin1]{inputenc}
\usepackage{hyperref}
\usepackage{textcomp}
\usepackage[T1]{fontenc}
\usepackage{mathptmx}
\usepackage{color}

\begin{document}

%\title{Crosstalk of propagating and localized spin-wave modes}
%\title{Relaxation of spin waves in nickel after femtosecond laser pulse excitation}
\title{Unusual spin-wave population in nickel after femtosecond laser pulse excitation}

\author{Benjamin~Lenk}
\email{blenk@gwdg.de}
%\author{Henning~Ulrichs}
%\author{Jakob~Walowski}
\author{Gerrit~Eilers}
\affiliation{I.~Institute of Physics, University of G\"{o}ttingen, 37077~G\"{o}ttingen, Germany}
\author{Jaroslav~Hamrle}
\affiliation{Centre for Advanced Innovation Technology, VSB -- Technical University of Ostrava
708 33 Ostrava-Poruba, Czech Republic}
\author{Markus~M\"{u}nzenberg}
%\email{mmuenze@gwdg.de}
\affiliation{I.~Institute of Physics, University of G\"{o}ttingen, 37077~G\"{o}ttingen, Germany}

%\date{\today , \thistime}

\begin{abstract}
The spin-wave relaxation mechanisms after intense laser excitation in ferromagnetic nickel films are investigated with all-optical pump-probe experiments. Uniform precession (Kittel mode), Damon-Eshbach surface modes and perpendicular standing spin waves can be identified by their dispersion~$\omega(H_\mathrm{ext})$. However, different to other ferromagnets $\omega(H_\mathrm{ext})$ deviates from the expected behavior. Namely, a mode discontinuity is observed, that can be attributed to a non-linear process. Above a critical field the power spectrum reveals a redistribution of the energy within the spin-wave spectrum populated.
\end{abstract}



\maketitle

\section{Introduction}
Spin-wave generation and manipulation have already been demonstrated in many ways.\cite{Demokritov2004} They are of great importance in magnetism-based spin-wave logic and XNOR as well as NAND gates have recently been implemented.\cite{Schneider08}
To reduce the size of potential devices, effective spin-wave pumping mechanisms are needed in ferromagnetic metallic films owning micron to nanometer characteristic magnetic length scales.
Here we present a pumping mechanism using fs-laser excitation that allows for a large precession angle ($>3^\circ$) after local excitation.
On the other hand, the relevant length scales can be artificially selected by the design of magnonic crystals, i.e.\ periodically micro- and nano-structured magnetic materials, with only very selected dynamic eigen states.

For large precession angles the equations of motion are intrinsically non-linear. This can be used to pump energy into a certain spin-wave mode by non-linear interaction and becomes evident for example in spin-wave nano-oscillators.\cite{Slavin06-locking,Kaka05-locking,Mancoff05-locking}
Given a spatial separation of two spin-torque oscillators and hence, an overlap in their spin-wave power when excited by a direct current, one finds a locking of frequencies if the frequency mismatch is below a critical value.
Driven to an extreme, a steady transfer of energy from higher to lower energy modes results in Bose-Einstein condensation observed at room temperature.\cite{Demidov2008}

% FIGURE 1
%
\begin{figure}
\includegraphics[width=0.99\linewidth]{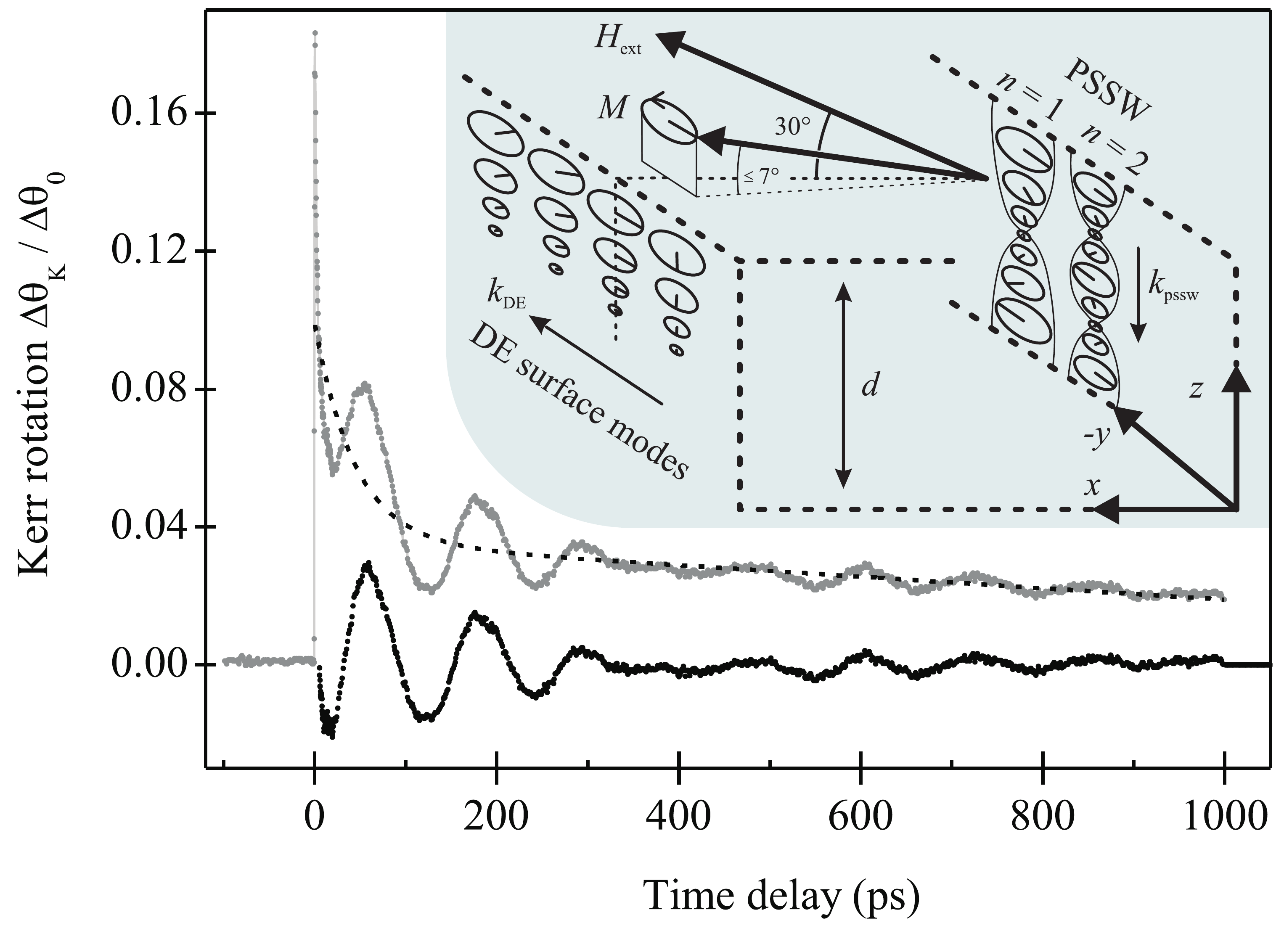}
\caption{Time-resolved measurement of the pump-induced change of the Kerr rotation (light gray) and the fitted background (dashed line) that is subtracted to yield the residual oscillation (black). The data set shown has been recorded at a nickel thickness of $80\,\mathrm{nm}$ under an external field of $140\,\mathrm{mT}$. A large angle of precession as high as $3^\circ$ is observed. Inset: Schematic representation of spin-wave modes in a continuous ferromagnetic film of thickness~$d$. Local amplitude and phase of the dynamic magnetization~$m(t)$ are illustrated. Shape anisotropy accounts for mainly in-plane alignment of the magnetization~$M = M_\mathrm{S} + m(t)$.}
\label{fig:scheme}
\end{figure}
In the following we focus on dipolar and exchange spin-wave modes in continuous films excited by intense laser pulses, where the nature of the excited spin-wave modes is tuned by the external magnetic field.
The excitation by femtosecond laser pulses has major differences as compared to methods that work in thermal equilibrium.
These are for example Brillouin light scattering (BLS),\cite{Jorzick99} conventional ferromagnetic resonance (FMR), strip-line based techniques like vector network analyzer (VNA)\cite{Neusser2008} or spatio-temporal Kerr effect based FMR used to study spin-wave modes.\cite{PechanJAP2005}
Our technique works in space and time-resolved manner and the detection process owns no $k$-selectivity a priori.
Yet, the strongest difference to resonant techniques is the broadband excitation in the spin system,\cite{marija07,Atxitia_2010_PRB} by which all magnetic modes are excited.
However, only resonant modes contribute to the signal coherently and can be detected while others appear as non-coherent background.
A further advantage is the contact-free, very local excitation and detection which simplifies the testing of various samples and structured media.
%The excitation imposes a major difference to other experimental approaches which operate at thermal equilibrium, e.g.\ Brillouin light scattering (BLS),\cite{Jorzick99} vector network analysis (VNA)\cite{Neusser2008} or ferromagnetic resonance (FMR).\cite{PechanJAP2005}
%More specifically, no phase information is carried by the pump pulses and spin-wave population takes place from a highly disordered initial state.

We make the very unusual observation, that below a critical magnetic field two dominant spin-wave modes are excited, whereas above the critical field, they merge into one single mode. %Even though not spatially separated, the modes behave very similar to the nano-oscillators described above, which can be regarded as a model system.
We shall also see, that intrinsic to the experimental method used, surface modes are preferentially excited if the sample thickness considerably exceeds the penetration depth of the laser field.
%
% FIGURE 2
\begin{figure*}
\includegraphics[width=\textwidth]{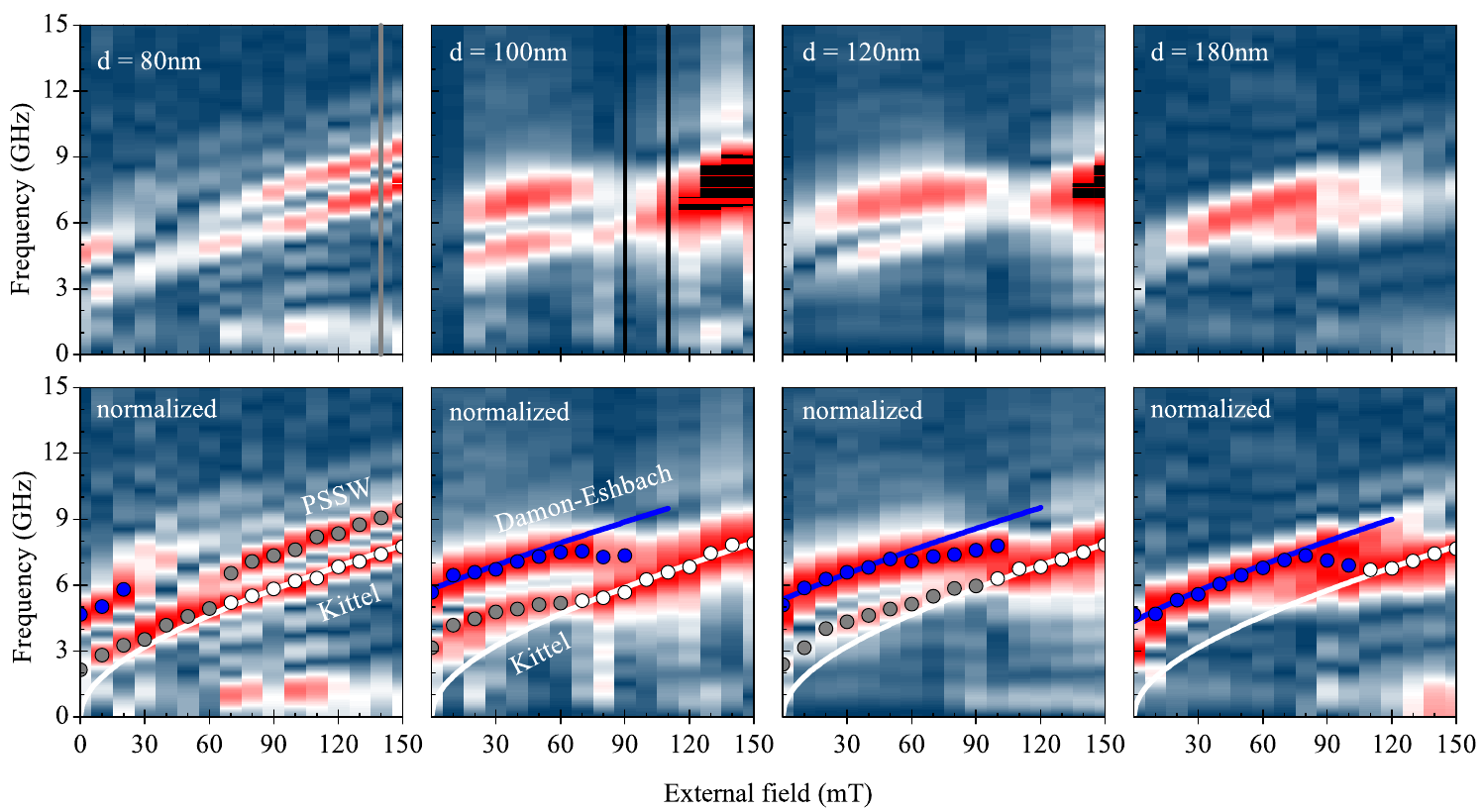}
\caption{(color online). Spin-wave power spectra of thin nickel films with $d=80$-$180\,\mathrm{nm}$. Magnetic oscillations are observed after optical excitation (top plots) and normalization yields frequency branches (bottom plots), that can be attributed to uniform precession (white points), magnetostatic surface waves (blue points) as well as perpendicular standing spin waves (gray points). Solid lines represent theory according to Eqs.~\eqref{eq:kittel} to~\eqref{eq:pssw}. The vertical gray line at $80\,\mathrm{nm}$ of nickel corresponds to the data set from \autoref{fig:scheme}, vertical black lines for $100\,\mathrm{nm}$ denote Fourier spectra analyzed in \autoref{fig:03-fft}. Please note the mode discontinuity around $\mu_0 H_\mathrm{ext}=90\,\mathrm{mT}$ for $d>80\,\mathrm{nm}$.}
\label{fig:02-thickness}
\end{figure*}

In general, magnetic precessional motion can have dipolar and exchange spin-wave character and can be described as follows. Starting from the free magnetic energy density, the Landau-Lifshitz equation of motion for the magnetization vector~$M$ yields a theoretical dispersion describing the uniform precession (also referred to as the Kittel mode) in the macrospin approximation\cite{Farle98}
\begin{equation}\label{eq:kittel}
\left(\frac{\omega_\mathrm{k}}{\gamma\mu_0}\right)^2=H_x \left(H_x + M_\mathrm{S} - \frac{2K_z}{\mu_0M_\mathrm{S}}\right),
\end{equation}
where $\mu_0 M_\mathrm{S}$ is the saturation magnetization (being $=659\,\mathrm{mT}$ for nickel), the only free parameter $K_z$ accounts for the effective anisotropy in the out-of-plane direction, and $H_x$ is the component of the external field projected to the sample plane.
Damon and Eshbach solved the equation of motion accounting for Maxwell's equations in the magnetostatic limit and adequate boundary conditions yielding the dispersion for long wavelength surface modes, so called Damon-Eshbach (DE) modes:
\begin{equation}\label{eq:de}
\left(\frac{\omega_\mathrm{DE}}{\gamma\mu_0}\right)^2= H_x \left(H_x + M_\mathrm{S}\right)+\frac{M_\mathrm{S}^2}{4}\left[1-\exp(-2k_\mathrm{DE}d)\right].
\end{equation}
Here, $d$ is the film thickness and $k_\mathrm{DE}\perp M_\mathrm{S}$ denotes the wave vector of the magnetostatic surface wave. An analogous approach in Ref.~[\onlinecite{kalinikos86}] yields dispersion characteristics taking into account exchange interaction as well as anisotropy.
In thin magnetic layers one finds a quantization of the wave vector perpendicular to the sample plane and in analogy to Eq.~\eqref{eq:kittel} one can give a dispersion describing the exchange-dominated perpendicular standing spin waves (PSSW):
\begin{multline}\label{eq:pssw}
\left(\frac{\omega_\mathrm{pssw}}{\gamma\mu_0}\right)^2=\left(H_x + \frac{2A}{\mu_0 M_\mathrm{S}}\, k_\mathrm{pssw}^2\right)\times\\ \left(H_x + M_\mathrm{S} - \frac{2K_z}{\mu_0M_\mathrm{S}} + \frac{2A}{\mu_0 M_\mathrm{S}}\, k_\mathrm{pssw}^2\right).
\end{multline}
Therein, $A$ is the exchange constant and $k_\mathrm{pssw}=n\pi\, d^{-1}$ is the quantized wave vector in out-of-plane (i.e.\ $z$)-direction attributed to a given order~$n$ of the PSSW. A schematic of the different magnetic modes is given in the inset of \autoref{fig:scheme} where the Kittel mode is not explicitly drawn, as both PSSW and DE modes result in the uniform precession in the limit $k_\mathrm{pssw}\to 0$ (i.e.\ $n=0$) and $k_\mathrm{DE}\to 0$, respectively.

%The two different magnetic materials used are Nickel and Cobalt-Iron-Boron (CoFeB). The intrinsic, Gilbert-type, damping differs strongly ($\alpha_{\rm Ni}\approx xxx$ and $\alpha_{\rm CoFeB}\approx xxx$, respectively) resulting in

\section{Experiment}
The samples consist of polycrystalline nickel films with thicknesses of up to $220\,\mathrm{nm}$. The thickness dependence of the occurring modes is measured on a wedge-shaped sample to perform all measurements on one specimen. It was prepared by electron beam evaporation in ultra high vacuum under a base pressure of $5\times 10^{-10}\,\mathrm{mbar}$. A linearly moving shutter was used to produce a nickel wedge with thickness $20\,\mathrm{nm} \leq d \leq 220\,\mathrm{nm}$ on a Si(100) substrate and in order to prevent oxidation, the ferromagnetic layer was capped with $2\,\mathrm{nm}$ of copper.

For the experiments we use an approach in the time domain, i.e.\ an all-optical pump-probe setup as previously described in Reference~[\onlinecite{Marija06JAP}]. An ultrashort laser pulse with $60\,\mathrm{fs}$ duration and $\lambda_c=810\,\mathrm{nm}$ central wavelength excites the saturated sample and at a variable time delay~$\tau$ a second laser pulse with the same characteristics but $5\%$ of the intensity probes the time-resolved magneto-optical Kerr effect (TRMOKE).\cite{beau96,Ju1998,Kampen2002}
Therewith, spin dynamics in the femtosecond regime can be detected, where electronic excitations lead to a demagnetization and subsequent relaxation into the initial state via scattering of high energy to low energy spin waves.\cite{marija07}
The processes that initially quench the magnetization in the first few picoseconds are subject to ongoing discussions,\cite{Koopmans2010} but can be modeled quite successfully in some detail.\cite{Atxitia_2010_PRB}
Here in focus are oscillations on a timescale from $30\,\mathrm{ps}$ to $1\,\mathrm{ns}$. We shall see, that all modes corresponding to equations~\eqref{eq:kittel} to~\eqref{eq:pssw} can be optically excited in a thin ferromagnetic film.
The responsible mechanism can be thought as an effective field pulse which stems from the heat-induced change of the sample's anisotropy and local exchange field upon absorption of the pump pulse.\cite{liu:JAP:2007}
Therefore, to promote magnetic precession the external field was tilted by $30^\circ$ out-of-plane thus giving rise to an angle between sample plane and the effective internal field~$H_i\parallel M_\mathrm{S}$.
For an external field of $\mu_0 H_{\mathrm{ext}}=150\,\mathrm{mT}$, which is the maximum used in the experiments, we calculated a rotation of $7^\circ$ of $H_{i}$ out-of-plane due to the Zeeman term in balance with the shape anisotropy (see also \autoref{fig:scheme}).
The angle is rather small and thus, its influence can be neglected for simplicity. Hence, in the analytical expression of the spin-waves modes we assume $H_i$ and $M_\mathrm{S}$ to be in-plane.

% FIGURE 3
%
\begin{figure}%
\centering%
\includegraphics[width=0.99\linewidth]{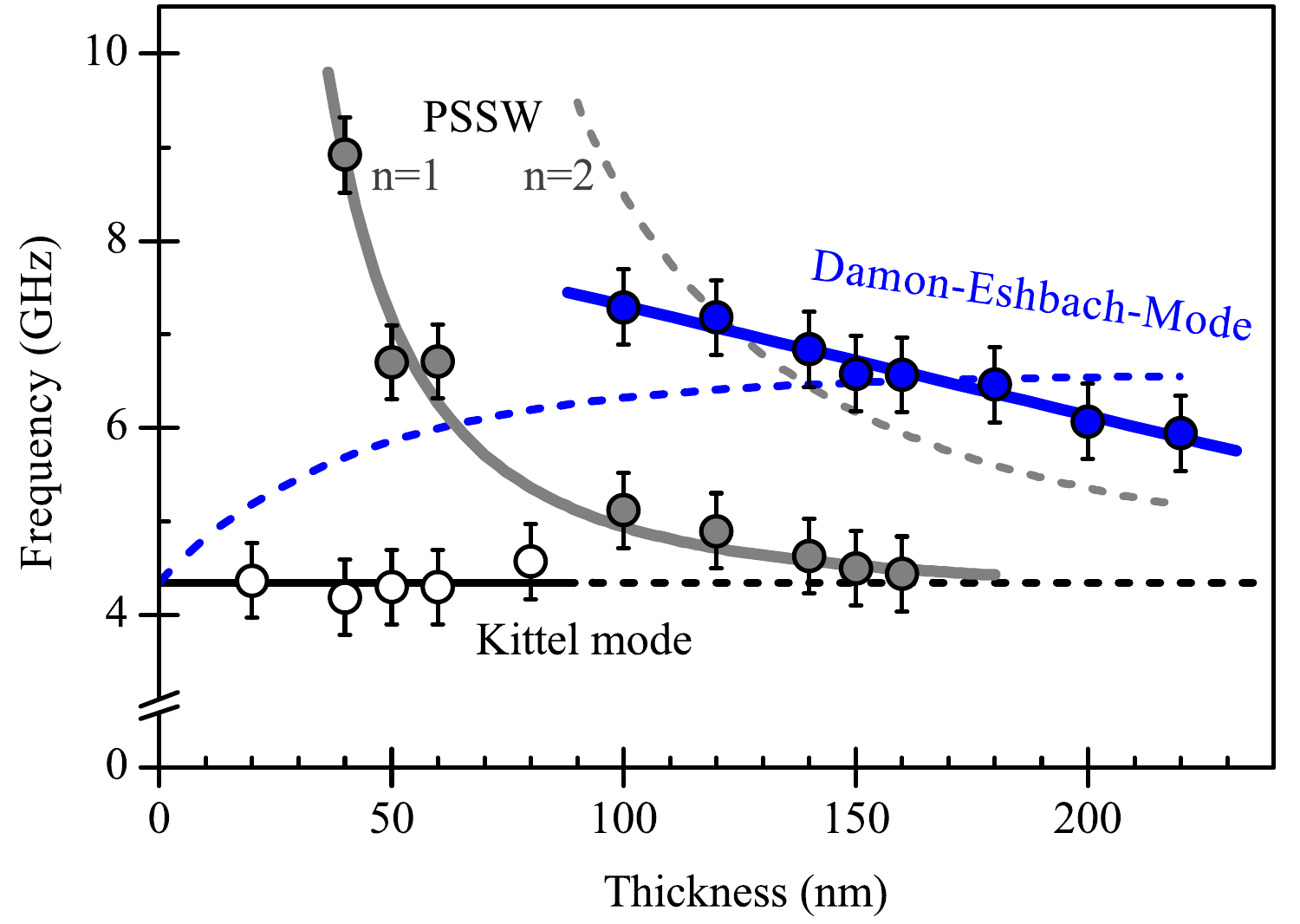}
\caption{Spin-wave modes at  $\mu_0 H_{\mathrm{ext}}=50\,\mathrm{mT}$ for varying thickness~$d$. Below $40\,\mathrm{nm}$ only the uniform precession (white points, black line) is observed, which does not change in frequency for increasing~$d$. Perpendicular standing spin waves with $n=1$ can also be identified (gray) and fit well to the expected $d^{-2}$ behavior.\cite{Seavey1958} PSSW of second order are not observed (dashed gray line), instead Damon-Eshbach surface waves (blue) are excited. The blue dashed line represents the DE dispersion~\eqref{eq:de} with $k_\mathrm{DE}=1\,\text{\textmu m}^{-1}$ and reveals that the DE wave vector is not constant in our experiment (the solid blue line is a guide to the eye).}%
\label{fig:freqs}%
\end{figure}%
\section{Results and discussion}
In order to obtain the precession frequencies from the time-dependent MOKE spectra, the incoherent background owing to phonons and magnons is subtracted. This procedure is demonstrated in \autoref{fig:scheme} with a single measurement recorded at $\mu_0 H_\mathrm{ext}=140\,\mathrm{mT}$ and $d=80\,\mathrm{nm}$.
After subtraction of the background (black data points in \autoref{fig:scheme}), Fourier transformation of the $M(\tau)$ curves yields peaks in the oscillations' power spectra. For a given thickness~$d$, the external field is varied between $0$ and $150\,\mathrm{mT}$, resulting in a change of amplitude and frequency of the oscillation.
The position of the corresponding peaks in the frequency domain is determined for each such set of measurements and plotted versus~$\mu_0 H_\mathrm{ext}$ (for details of the analysis refer to the supplementary data).

% FIGURE 4
%
\begin{figure}
\includegraphics[width=0.99\linewidth]{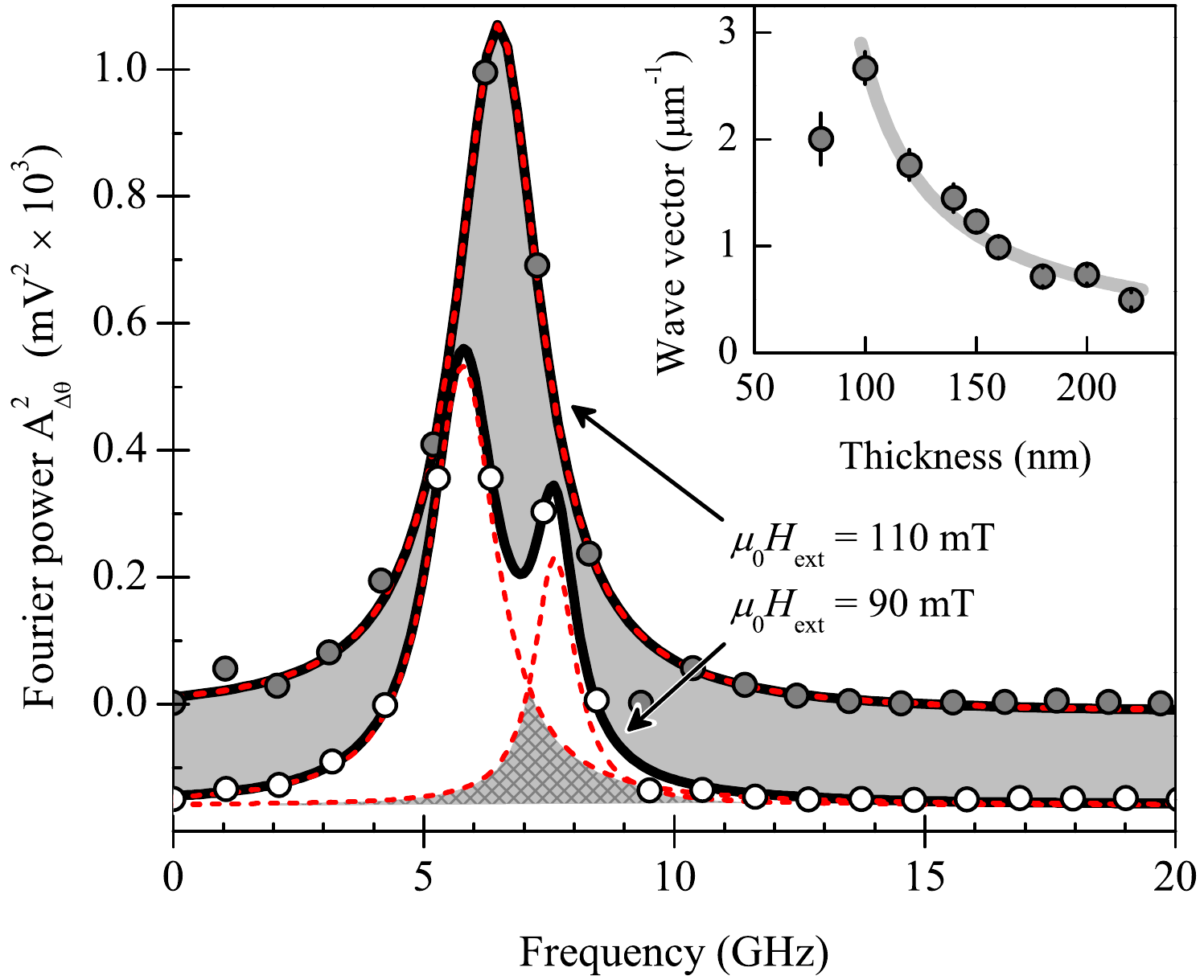}
\caption{(color online). Gray and white points denote the Fourier power~$A^2_{\Delta\theta}$ as calculated from the time-resolved MOKE measurements at $d=100\,\mathrm{nm}$, $\mu_0 H_{\mathrm{ext}}=110\,\mathrm{mT}$ and $90\,\mathrm{mT}$, respectively. Solid and dashed lines represent Lorentz-Peaks which where fitted to the data. The expected linewidth in nickel is larger than $1\,\mathrm{GHz}$ and as a result, possible mode locking leads to a single peak at $110\,\mathrm{mT}$ if the overlap indicated by the hatched area is increased. Additionally, the mutual intensity is strongly increased. In the inset, the dependence of the Damon-Eshbach wave vector on the thickness is given, where the solid gray line is a guide to the eye (see text also).}
\label{fig:03-fft}
\end{figure}
Shown in \autoref{fig:02-thickness} are power spectra recorded on different nickel thicknesses, revealing up to three precessional modes of different origin.
In the upper row the Fourier power spectra for a given thickness are plotted in a color map as calculated from the $M(\tau)$ curves (the Fourier transformation of the data from \autoref{fig:scheme} is marked with a vertical gray line).
In the bottom row all spectra have been normalized by their respective maximal FFT power to give a better overview of mode evolution. The data points represent the peak positions and included as solid lines are the fitted theoretical frequency dispersions \eqref{eq:kittel}-\eqref{eq:pssw} of the various modes, which have been used to attribute the branches to the Kittel mode, the DE surface mode as well as the PSSW mode with $n=1$.

% FIGURE 5
%
\begin{figure*}
\centering
\includegraphics[width=\textwidth]{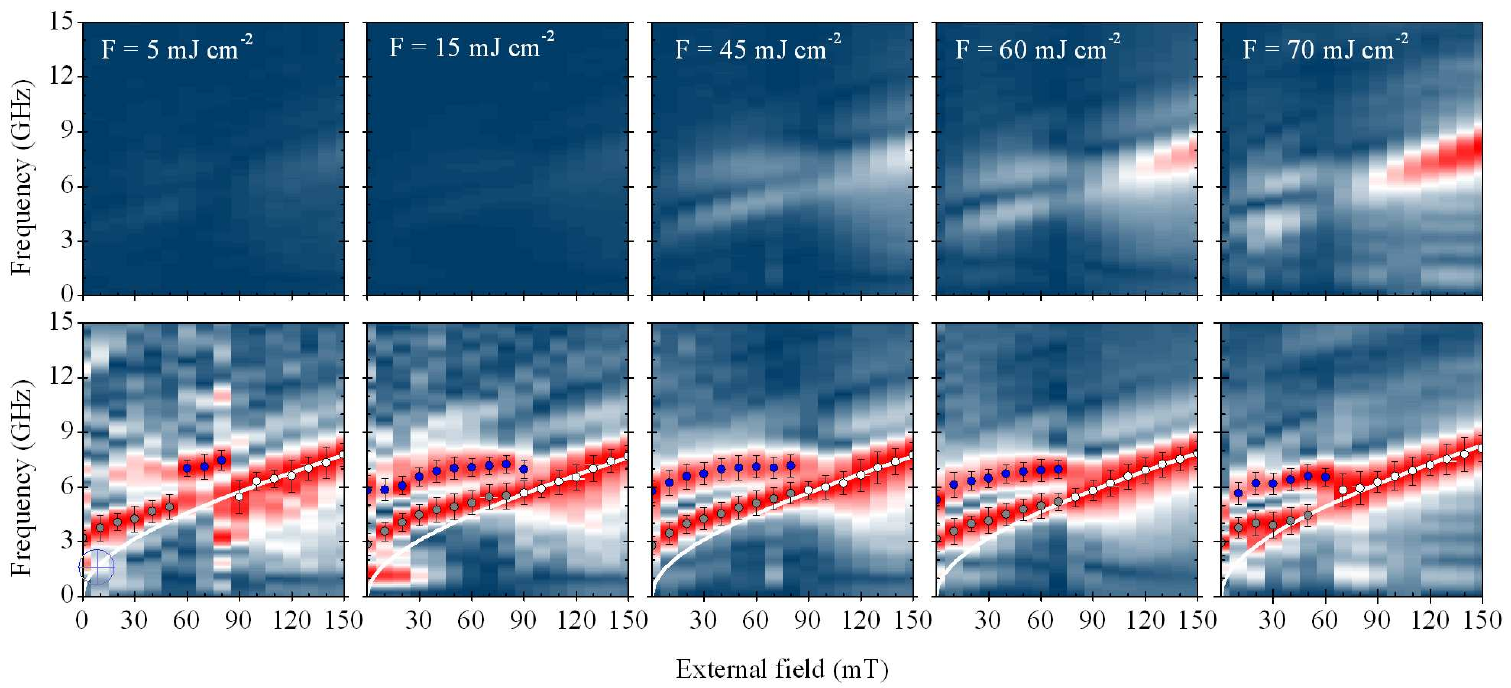}
\caption{(color online). Fourier power spectra for different pumping fluences $F=5$-$70\,\mathrm{mJ\, cm^{-2}}$ with $d=150\,\mathrm{nm}$. The identification of modes corresponds to \autoref{fig:02-thickness} and the color scale in the top row (not normalized plots) is the same for every spectrum to reveal the fluence-dependence of the precession amplitude. One observes a strongly increasing amplitude of the uniform precession with increasing fluence, whereas the bottom row (normalized data) shows a small shift of the critical field towards lower values, which will be detailed in \autoref{fig:05-population}.}
\label{fig:04-fluence}
\end{figure*}
Analysis of the thickness dependent occurance of the different magnetic modes yields the results depicted in \autoref{fig:freqs}. For a given external field of $50\,\mathrm{mT}$ the observed frequencies have been extracted and the aforementioned identification of Kittel-, PSSW-, and DE-mode is confirmed.
Both the Kittel mode (black) and the PSSW (gray) display the expected evolution with the tickness: While the frequency of uniform precession~$\omega_\mathrm{K}$ remains unchanged, the frequency of the PSSW reduces.
Being proportional to $d^{-2}$ it approaches $\omega_\mathrm{K}$ in accordance to literature.\cite{Seavey1958,footnote_pssw}

The DE modes (blue lines and points in \autoref{fig:freqs}) have been predominantly studied with Brillouin light scattering (BLS)\cite{Gruenberg82}. We find from our experiment that they are first excited at a nickel thickness of $80\,\mathrm{nm}$.
There, they are only present at low external fields, but for thicker nickel layers these modes dominate the spin-wave dynamics over a wider field range up to $\mu_0 H_\mathrm{ext}=90\,\mathrm{mT}$.
This is due to the excitation mechanism in combination with the relevant length scales in our experiment: sample thickness~$d$ and optical penetration depth~$\Lambda_\mathrm{opt}$ (i.e.\ penetration depth of the electrical field at optical frequencies which evaluates to $\lambda/(2\pi\Im(N))\approx 30\rm\, nm$, where $N$ is the complex refractivity index).
Calculations show that our experiment is sensitive to magnetic modes in the top $15\,\mathrm{nm}$ of the ferromagnet, corresponding details will be published elsewhere.\cite{hamrle}
The magnetic excitation depth is governed by the penetration of the laser field and energy is deposited in form of an initial disarrangement of spins.\cite{marija07}
This energy will very effectively excite eigen modes with similar spatial profile and thus, by choosing a thickness~$d\gg\Lambda_\mathrm{opt}$, the resulting strong asymmetry will make DE surface modes the favored relaxation channel.
Consequently, the appearance of the DE modes is based on the fact that they own an amplitude profile with the maximum at the surface, decaying exponentially into the film.
Herein lies a strong analogy to surface acoustic waves excited with laser pulses, which can be used to determine elastic properties of surface layers.\cite{Neubrand1992}

Assuming a propagation direction perpendicular to the magnetization ($k_\mathrm{DE}\perp M_\mathrm{S}, H_\mathrm{ext}$) the dispersion~\eqref{eq:de} can be used to determine the wave vector~$k_\mathrm{DE}$.
In the inset of \autoref{fig:03-fft} the respective quantitative analysis of the Damon-Eshbach modes is given.
From the fits one obtains wave vectors in the inverse micrometer range, corresponding to wave lengths of about $3\,\text{\textmu m}$. We find that the product $k_\mathrm{DE}d$ entering Eq.~\eqref{eq:de} is not constant but that the relation $k_\mathrm{DE}=\left(d-d_0\right)^{-1}$ holds with $d_0=66(4)\,\mathrm{nm}$ (solid gray line in the inset of \autoref{fig:03-fft}).
The divergence of $k_\mathrm{DE}$ around that value supports our earlier interpretation.
An asymmetry between film thickness and optical excitation depth is needed to give rise to dipolar spin waves in contrast to exchange-dominated spin waves with rather high~$k$.
In other words, the parameter $d_0$ can be considered an onset thickness above which the dipolar interaction comes into play.
In general, our understanding of the selection of the DE wave vector is not fully developed. However, we can exclude, that the laser spot sizes of pump ($60\,\text{\textmu m}$) or probe ($20\,\text{\textmu m}$) enforce a selection, since the observed DE wave lengths are of the order of only very few microns and we do not observe any dependencies on the pump or probe beam diameter.

As described, the DE mode dominates the spectra up to critical fields as high as $\mu_0 H_\mathrm{crit}=90\,\mathrm{mT}$. Around that value, an explicit deviation from the theoretical dispersion arises (blue line and points in the bottom row of \autoref{fig:02-thickness}).
A distinct modification of the DE mode intensity takes place which --~for sufficiently high fields~-- ultimately results in a peculiar characteristic evident in \autoref{fig:02-thickness}. Namely the DE mode merges into the Kittel mode.
This leads to a very strong increase of the total Fourier power, i.e.\ precession amplitude, as seen in the non-normalized data (upper row of \autoref{fig:02-thickness}). The propagating surface mode seems to lock to the frequency of the Kittel and PSSW modes for magnetic fields higher than $\mu_0 H_\mathrm{crit}$.
Above that value, the peaks can be attributed to the dispersion of the uniform precession by Eq.~\ref{eq:kittel}. The fits (white lines) yield values for the effective anisotropy~$K_\mathrm{z}$ around $50\, \mathrm{kJ\, m^{-3}}$ showing no monotonous trend with the thickness.

To discuss this unusual observation we want to remind the reader, that the highly asymmetric initial excitation profile unfavors the uniform precession as a magnetic eigen mode.
One possibility to understand its observation is a locking of the DE mode to the $k=0$ homogenious precession originating from an energy transfer between the two dynamic modes.
Only if the propagating surface waves would couple to the localized Kittel mode the amplitude of the magnetic precession could be that drastically increased.

A reasonable starting point to discuss this idea is to draw parallelities to a model by Slavin and coworkers. They proved, that phase locking of spin-torque nano-oscillators can take place.\cite{Slavin06-locking}
In their publication, the authors use a non-linear set of equations of motion for two coupled oscillators and can explain the experimental results presented in Refs.~[\onlinecite{Kaka05-locking,Mancoff05-locking}].
Once the free running frequency mismatch between the two oscillators is smaller than a threshold value~$\Delta_\mathrm{max}$ (Eq.~(9) in Ref.~[\onlinecite{Slavin06-locking}]), they phase lock to a mutual frequency. The mechanism is twofold: coupling can be either mediated by spin waves radiated into a common magnetic layer or a dipole field created by the oscillators.
Either way, one of the crucial parameters is the spatial separation~$a$ of the two contacts, tuning the overlap of spin-wave power in real space (Fig.~1 in Ref.~[\onlinecite{Slavin06-locking}]).

Transferring these findings to our observation, we could speculate that an increased overlap of the magnetic modes in Fourier space leads to a sudden interaction such that locking takes place. Figure~\ref{fig:03-fft} illustrates this hypothesis: two Fourier spectra from measurements recorded on nickel with a thickness of $100\,\mathrm{nm}$ are shown.
At $\mu_0 H_\mathrm{ext}=90\,\mathrm{mT}$ clearly two precessional modes can be distinguished, whereas at $\mu_0 H_\mathrm{ext}=110\,\mathrm{mT}$ only one smooth peak with nearly identical width is found.
The two modes observed, namely uniform precession and DE surface waves, are separated in Fourier space by several GHz at low external fields. However, if $\mu_0 H_\mathrm{ext}$ is increased in the experiment, the separation reduces and at one point approaches the value of the linewidth, which implies an overlap in the frequency domain (hatched area in \autoref{fig:03-fft}).
Above a threshold overlap the frequencies seem to lock and the two modes cannot be distinguished anymore. Moreover, a single smooth peak in the Fourier spectrum is observed.
Concerning the evolution with increasing magnetic field, a step-like change of the DE frequency is apparent in the Fourier spectra in \autoref{fig:02-thickness} and hence, the term mode locking seems justified.

Equivalent to the damping time constant $\tau_\alpha$ in the time domain is the linewidth in the Fourier domain. From $\tau_\alpha$ apparent in the time-resolved MOKE spectra one expects the width of the peak after Fourier transformation to be $(\pi\tau_\alpha)^{-1}$ at 50\% of maximum power (FWHM, full width half maximum).
For nickel the expected linewidth therefore is $1.1\,\mathrm{GHz}$, using $\tau_\alpha= 300\,\mathrm{ps}$, which has been extracted from the time-resolved spectra.
Due to the strong damping in nickel (implying a rather large line width) considerable overlap can occur. The interplay with $H_\mathrm{ext}$ tuning the DE and Kittel frequency mismatch leads to possible phase locking (see \autoref{fig:03-fft}).

% FIGURE 6
%
\begin{figure}
\centering
\includegraphics[width=0.99\linewidth]{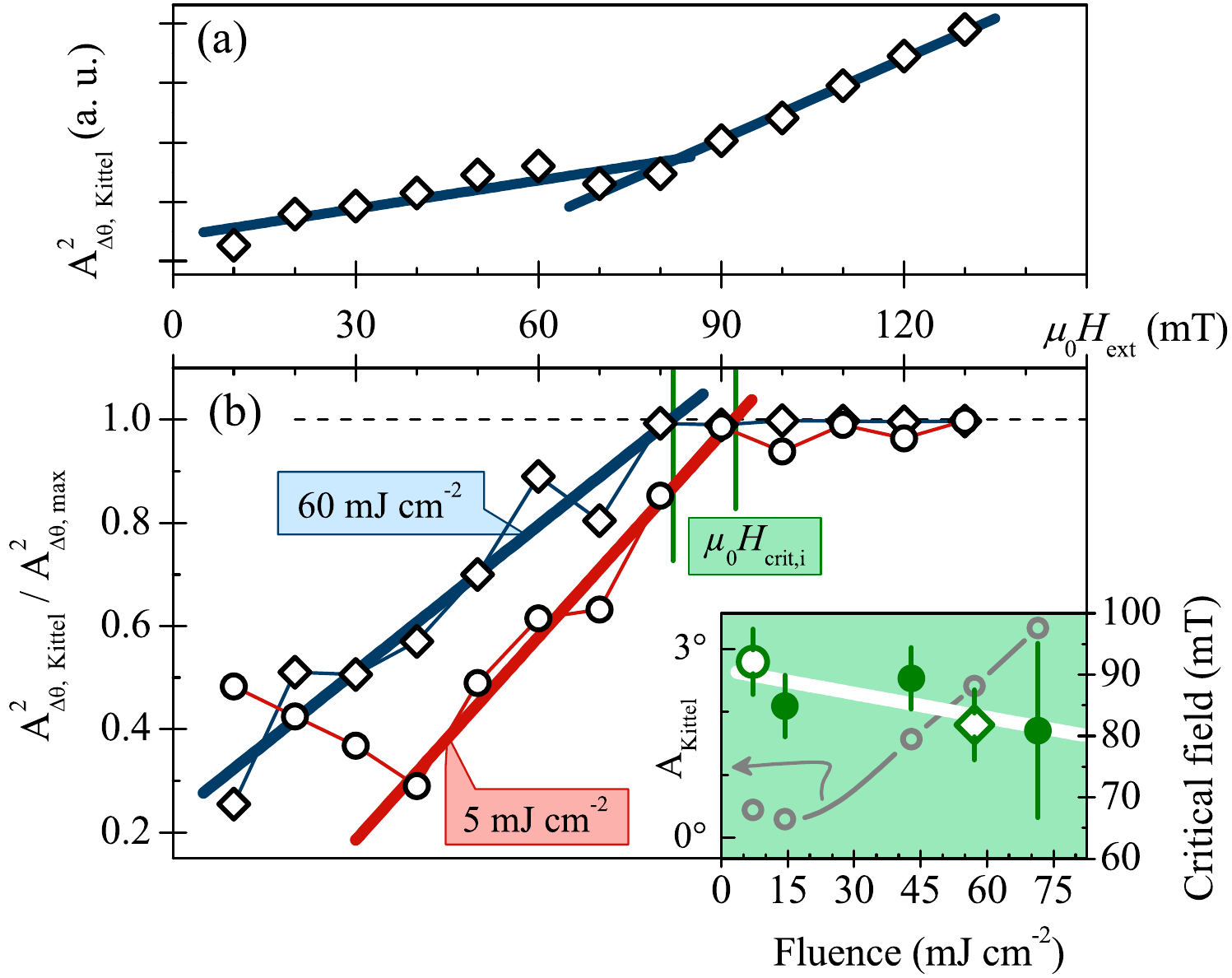}
\caption{(color online). (a) Threshold behavior of the Fourier power~$A^2_{\Delta\theta,\mathrm{Kittel}}$ of the Kittel mode. The external magnetic field can be used tune the observed characteristics between dipolar surface waves and uniform precession. (b) Determination of the critical magnetic field~$\mu_0 H_\mathrm{crit}$ for different fluences~$F = 5$ and $60\,\mathrm{mJ\, cm^{-2}}$, respectively. $A^2_{\Delta\theta,\mathrm{Kittel}}$ is plotted after normalization to the maximum of the respective Fourier spectrum for a given external field,~$A^2_{\Delta\theta,\mathrm{max}}$. The result can be regarded as a measure of mode population and a linear increase is found prior to saturation. The critical field can be extracted as depicted by the vertical green lines. In the inset, corresponding results for all fluences are included, giving a clear trend towards smaller $\mu_0 H_\mathrm{crit}$ for increasing~$F$. Additionally, the Kittel amplitude as induced by different pump fluences is shown for $\mu_0 H_\mathrm{ext}=150\,\mathrm{mT}$ (gray points).}
\label{fig:05-population}
\end{figure}
A test for this hypothesis based on non-linearity is the influence of the (optical) pumping power that excites the magnetization dynamics. Corresponding experiments were performed on a separate continuous nickel film which was $150\,\mathrm{nm}$ thick and was prepared as described above.
For the measurements, the pumping fluence was varied over one decade ranging from $5$ to $70\,\mathrm{mJ\, cm^{-2}}$.
The resulting Fourier power spectra are presented in \autoref{fig:04-fluence} and reveal that the nature of the magnetic modes excited by the pump pulses does not significantly change. To give a better insight into the fluence dependence, all power spectra in the top row are identically color-scaled.
This shows an increase of the spin-wave amplitude with the fluence (from left to right) as expected when considering the heat-induced disorder triggering the oscillation.\cite{liu:JAP:2007}
This increase is quantitatively given in the inset of \autoref{fig:05-population}(b), where the Kittel amplitude at $\mu_0 H_\mathrm{ext}=150\,\mathrm{mT}$ has been plotted versus the pumping fluence (gray points).
In this respect, the absolute value of the amplitude is of great interest. It can be derived from the dynamic magnetization by calibration of the experiment for a given saturation magnetization:
A comparison of the magneto-optical signal at zero time delay with and without pump-induced demagnetization at $F=70\,\mathrm{mJ\, cm^{-2}}$ and $\mu_0 H_\mathrm{ext}=150\,\mathrm{mT}$ results in a precessional amplitude of $4.6\% \times \mu_0 M_\mathrm{S}= 30\,\mathrm{mT}$, corresponding to an angle of~$3^\circ$.

A feature of \autoref{fig:04-fluence} that requires a closer look is the somewhat changing critical field for the transition from surface to uniform characteristics. Figure~\ref{fig:05-population}(b) details the corresponding analysis and gives the respective results in the inset.
Starting from the fit of the Kittel dispersion~\eqref{eq:kittel} to the data (white lines in the bottom row of \autoref{fig:04-fluence}) the amplitude of the Kittel mode for each magnetic field is extracted from the data.
In the normalized case, this amplitude equals $1$ if $H_\mathrm{ext}>H_\mathrm{crit}$ and below the critical field a monotonous, in good approximation linear behavior is found.
By a linear fit, the point at which the Kittel amplitude reaches $1$, i.e.\ above which the Kittel mode is the dominant mode of precession can be determined.
We define this field value as the critical field~$H_\mathrm{crit}$ (also marked by vertical lines in \autoref{fig:05-population}(b)). The above described procedure is explicitly given in \autoref{fig:05-population}(b) for the fluences $5$ and $60\,\mathrm{mJ\, cm^{-2}}$, respectively.
In the inset the overall decrease of $\mu_0 H_\mathrm{crit}$ for increasing $F$ is shown.
Together with the threshold-like onset of the uniform precession in \autoref{fig:05-population}(a) and the accompanying rapid increase of precession amplitude our initial hypothesis of a non-linear transfer of energy seems to be verified.

\section{Conclusion}
In conclusion, we have identified the excitation of the uniform precession (Kittel mode) as well as magnetostatic surface waves (DE mode) in asymmetrically pumped nickel films.
Below a critical magnetic field~$H_\mathrm{crit}$ the dipolar DE modes dominate the precession, whereas a step-like feature in the dispersion~$\omega_\mathrm{DE}(H_\mathrm{ext})$ indicates the onset of the population of the $k=0$ mode above $H_\mathrm{crit}$ (\autoref{fig:02-thickness}).
The observation of the DE surface waves for $H_\mathrm{ext}<H_\mathrm{crit}$ can be well understood by considering the optical excitation mechanism.
A concept was developed taking into account the amplitude profile of the DE modes as well as the asymmetric excitation profile stemming from the laser pump pulses.
However, the understanding of the selection of relaxation channels in terms of the selection of the DE wave vector is a challenge still to be met.
The transistion to uniform precession for $H_\mathrm{ext}>H_\mathrm{crit}$ was discussed in terms of a non-linear mechanism transferring energy between the two eigen modes of the ferromagnet. Only a non-linear process allows to explain the dependence of the transition field~$H_\mathrm{crit}$ on the pumping power.
In this context the high energy of the pump pulses driving the system far out of equilibrium leading to large precession angles should be mentioned.
In total, the population of either the Kittel or the DE mode can be tuned by both the external field and the pumping fluence. Understanding the underlying mechanism might prove crucial on the way to spin-logic devices, especially with the interconversion of electrical, optical or magnetic signals in mind.

One of the authors (J.H.) acknowledges financial support through Grant Academy of The Academy of Sciences of the Czech Republic (KAN400100653).
Also, fruitful discussions with A.\ Slavin and B.\ Hillebrands are gratefully acknowledged.


\begin{thebibliography}{24}
\bibitem[{\citenamefont{footnote}()}]{footnote_pssw}
\bibinfo{author}\bibnamefont{From the fit in \autoref{fig:freqs} (solid gray line) the
  exchange constant~$A=6.9(5)$~pJ~m$^{-1}$ was extracted.}.
\end{thebibliography}
\end{document}